\documentclass{aipproc}

\newcommand{\z}{&&\hspace*{-0.7cm}}
\newcommand{\zz}{&&\hspace*{-0.3cm}}

\usepackage{epsfig}

\usepackage{longtable} % load only if you really need it

\begin{document}

\title{The differential equation method:\\
evaluation of complicated Feynman integrals}
\author{A. V. Kotikov\thanks{In parts supported by 
Alexander von Humboldt fellowship and RFBR (98-02-16923)}}
\affiliation{Particle Physics Laboratory, Joint Institute for Nuclear 
Research, 141980 Dubna, Russia}

\begin{abstract}
% Oh\footnote{Who likes footnotes to the abstract?} 

We discuss a progress in calculation of Feynman integrals which has been
done with help of the differential equation method and 
demonstrate the results for a class of two-point two-loop diagrams.

\end{abstract}

\maketitle

%\section{Introduction}

The idea of the differential equation method (DEM) (see [1,2]): to apply
the integration by parts procedure [3] to an internal $n$-point subgraph
of a complicated Feynman diagram and later to represent new complicated
diagrams, obtained here, as derivatives in respect of corresponding mass 
of the initial diagram.

Thus, we have got the differential equations for the initial diagram. The
inhomogeneous terms contain only more simpler diagrams.
These simpler diagrams have more trivial topological structure 
and/or less number of loops [1] and/or ends [2].

Applying the procedure several times, we will
able to represent complicated Feynman integrals (FI) and their derivatives
(in respect of internal masses) through a set of quite simple well-known 
diagrams. 
Then, the results for the
complicated FI can be obtained by integration several times
of the known results for corresponding simple diagrams.

Sometimes it is useful (see [4]) to use external momenta (or some their
functions) but not masses as parameters of integration.\\

%\newpage
%\vskip 0.5cm

{\bf The recent progress in calculation of Feynman integrals with help 
of the DEM.} 

{\bf 1.} The articles [5] and [6]: 

{\bf a)} The set of two-point two-loop FI with one- and two-mass
thresholds (see Fig.1) has been evaluated by 
a combination of DEM and Veretin programs
for calculation of first terms of FI small-moment expansion.
%, where solid lines denote propagators with the mass $m$; dashed lines
%denote massless propagators).
The results are given below
%on pages 3 and 4 
(some ot them have been known before (see disscussions in [5])). 
%The check of the results has been
%done by Veretin program.\\
%\smallskip
%%%  FIGURE  ===  %%%%%%%%%%%%%%%%
%   \begin{figure}[tb]
%%\label{fig-2}
%\unitlength=1mm
%\begin{picture}(150,100)
%  \put(0,100){%
%   \epsfig{file=tolya.ps,width=100mm,height=150mm,angle=-90}%
%}
%\end{picture}
%\vskip 0.5cm
% \caption{\sl  
%Two-loop selfenergy diagrams.
%Solid lines denote propagators with the mass $m$; dashed lines
%denote massless propagators.}
%\vskip 0.5cm
% \end{figure}

%Figure~\ref{fig:c} is an Escher graphic. 
%\smallsample 
\begin{figure}
%\resizebox{\columnwidth}{!}{\includegraphics{escher}}
\rotatebox{270}{
\resizebox{\columnwidth}{!}{\includegraphics{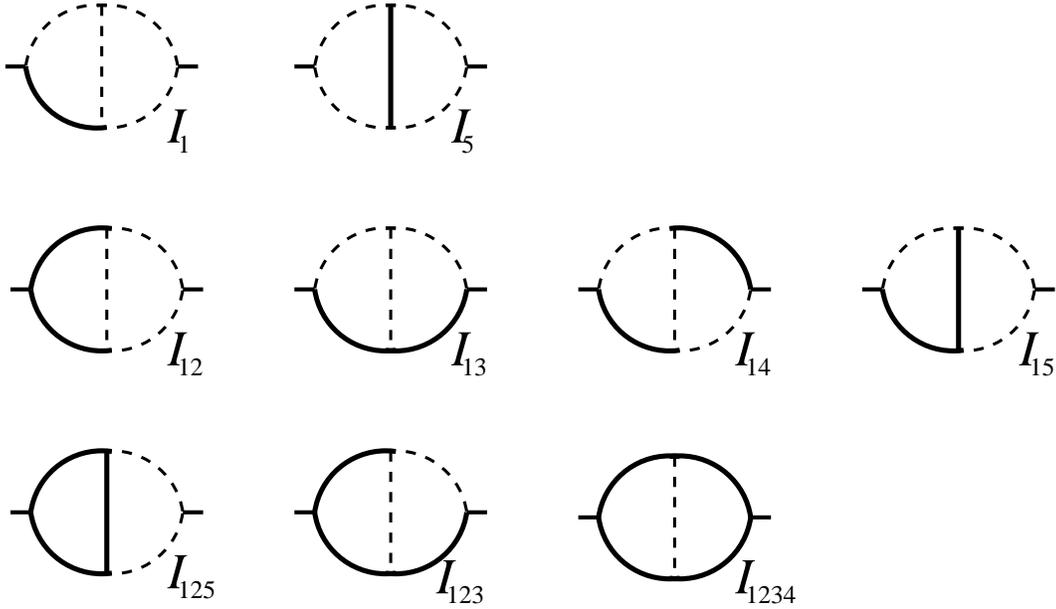}}
}
\caption{Two-loop selfenergy diagrams.
Solid lines denote propagators with the mass $m$; dashed lines
denote massless propagators.}
\label{fig:c}
\end{figure}
%\vskip -0.5cm

%\sample

{\bf b)} The set of three-point two-loop FI with one- and two-mass
thresholds has been evaluated (the results of some ot them has been
known before (see [6])) by a combination of DEM and Veretin programs
for calculation of first terms of FI small-moment expansion.
The finite parts of the integrals can be written in terms
of the generalized Nielsen polylogarithms (see [8]).

{\bf 2.} The article [7]: 

%The full set of two-point two-loop onshell master diagrams 
%has been evaluated by DEM. The check of the results has been
%done by Kalmykov program. 
The full set ot two-loop onshell master diagrams has been evaluated by
DEM and Kalmykov programs for calculation of first terms of FI 
small-moment expansion.
It has been observed that finite 
%(as $\varepsilon \to 0$) 
parts of all
such integrals without subdivergences can be written in terms of three
constants, two for the real part, $\zeta_3$ and
$\pi Cl_2(\pi/3)$, and one for the imaginary part, $\pi\zeta_2$.

{\bf 3.} The articles [9]: 

The set of three-point and four-point two-loop massless FI 
has been evaluated.\\

%XS\vskip 0.5cm
{\bf Here we demonstrate the results of FI are displayed on Fig.1.}
%only corresponding two-point FI and them 
%integral representations.}\\

We introduce the notation for
%\begin{equation}
%\label{Sqp}
%  S_{a+1,b}(z) = \frac{(-1)^{a+b}}{a!\,b!}
%    \int_0^1 \frac{\log^a(t)\log^b(1-zt)}{t}\,dt.
%\end{equation}
%
%  All 
polylogarithmic functions: 
%are particular cases of $S$ functions, namely
\begin{eqnarray}
%\label{Li}
\z  {\rm Li}_a(z) = S_{a-1,1}(z), ~~~ \nonumber \\
\z S_{a+1,b}(z) = \frac{(-1)^{a+b}}{a!\,b!}
   \int_0^1 \frac{\log^a(t)\log^b(1-zt)}{t}\,dt.
\nonumber 
\end{eqnarray}

  We introduce also the following two variables 
%($z=q^2/m^2$)
\begin{eqnarray}
z=\frac{q^2}{m^2},~~~y=\frac{1-\sqrt{z/(z-4)}}{1+\sqrt{z/(z-4)}}\,.
\nonumber
\end{eqnarray}

  Then
\begin{eqnarray}
\z q^2 \cdot I_1 = 
     - \frac{1}{2}\log^2(-z)\log(1-z) 
     - 2\log(-z){\rm Li}_2(z) \nonumber \\
 \zz + 3{\rm Li}_3(z) -6 S_{1,2}(z)  
%\nonumber \\   \zz- 
  -  \log(1-z) 
%\biggl(
[  \zeta_2 + 2{\rm Li}_2(z) ]
%\biggr)
\,, \nonumber \\
%
%\z\nonumber\\
%
\z q^2 \cdot I_5 = 
      2 \zeta_2 \log(1+z) + 2\log(-z){\rm Li}_2(-z) 
\nonumber \\
\zz+ \log^2(-z)\log(1+z)
        + 4\log(1+z){\rm Li}_2(z)  - 2{\rm Li}_3(-z) 
            \nonumber \\
      \zz 
           - 2{\rm Li}_3(z) 
           + 2 S_{1,2}(z^2) 
%\nonumber \\\zz
- 4S_{1,2}(z) - 4 S_{1,2}(-z)\,, \nonumber \\
%
%\z\nonumber\\
%
\z q^2 \cdot I_{12} =
     {\rm Li}_3(z) - 6 \zeta_3 - \zeta_2\log y
     - \frac16\log^3 y - 4\log y\,{\rm Li}_2(y)  \nonumber\\
     \zz+ 4{\rm Li}_3(y) - 3{\rm Li}_3(-y) + \frac13 {\rm Li}_3(-y^3)\,,
\nonumber \\
%
%\z\nonumber\\
%
\z q^2 \cdot I_{13} = 
      - 6 S_{1,2}(z)
      - 2 \log(1-z) 
%\biggl( 
[ \zeta_2 + {\rm Li}_2(z) ]
%\biggr)
\,, \nonumber \\
%
%\zz \nonumber\\
%
\z q^2 \cdot I_{14} =  
     \log(2-z) 
%\biggl( 
[\log^2(1-z) -2 \log(-z)\log(1-z) \nonumber \\
\zz- 2{\rm Li}_2(z)]
%       \biggr)
%\nonumber \\  \z-\z 
-\frac{2}{3} \log^3(1-z) + \log(-z)\log^2(1-z)
\nonumber \\
    \zz - 2 \zeta_2 \log(1-z) 
%\nonumber\\  \z-\z  
-S_{1,2}\bigr(1/(1-z)^2\bigl) + 2 S_{1,2}\bigr(1/(1-z)\bigl)
\nonumber \\
     \zz   + 
2 S_{1,2}\bigr(-1/(1-z)\bigl)  
%\nonumber\\  \z+\z 
+\frac13 \log^3 y 
+ \log^2 y\,
%\biggl( 
[2 \log(1+y^2)
\nonumber \\
   \zz -3 \log(1-y +y^2)]
% \biggr) 
- 6 \zeta_3 - {\rm Li}_3(-y^2)
     +\frac23 {\rm Li}_3(-y^3)
                   \nonumber \\
\zz - 6 {\rm Li}_3(-y) 
%         \nonumber \\ 
%\zz
+ 2\log y\, 
%\biggl( 
[{\rm Li}_2(-y^2) -{\rm Li}_2(-y^3) 
              +3 {\rm Li}_2(-y)    ]
%\biggr)
\,,  \nonumber \\
%
%\z\nonumber\\
%
\z q^2 \cdot I_{15} = 
   2{\rm Li}_3(z) - \log(-z)\,{\rm Li}_2(z) 
              + \zeta_2 \log(1-z) \nonumber \\
  \zz+ \frac16 \log^3 y 
   - \frac12\log^2 y\,
%\biggl( 
[8 \log(1-y) 
%\nonumber \\ \zz
- 3 \log(1-y +y^2) ]
%\biggr) 
                   \nonumber \\
\zz- 6 \zeta_3 - \frac13 {\rm Li}_3(-y^3) + 3 {\rm Li}_3(-y) 
            + 8 {\rm Li}_3(y) \nonumber \\ 
\zz+ \log y\, 
%\biggl( 
[{\rm Li}_2(-y^3) -3 {\rm Li}_2(-y) 
              -8 {\rm Li}_2(y)   ]
%\biggr)
\,, \nonumber \\
%
%\z\nonumber\\
%
\z q^2 \cdot I_{123} = 
    - \zeta_2 
%\biggl(  
[\log(1-z) +  \log y ]
%\biggr)
    - 6 \zeta_3  
\nonumber \\
\zz
- \frac32 \log(1-y +y^2)\, \log^2 y \nonumber \\
    \zz+ {\rm Li}_3(-y^3) - 9 {\rm Li}_3(-y)
%\nonumber \\   \zz
- 2 \log y\, 
%\biggl( 
[{\rm Li}_2(-y^3) - 3{\rm Li}_2(-y)]
% \biggr)  
      \,,\nonumber \\
%
%\z\nonumber\\
%
\z q^2 \cdot I_{125} =  
      -2 \log^2 y\, \log(1-y)
      - 6 \zeta_3 
%\nonumber \\
%\zz
+ 6 {\rm Li}_3(y) 
\nonumber \\
\zz
- 6\log y\, {\rm Li}_2(y)\,, \nonumber \\
%
%\z\nonumber\\
%
\z q^2 \cdot I_{1234} = 
    - 6 \zeta_3  - 12 {\rm Li}_3(y) - 24 {\rm Li}_3(-y)
  + 8 \log y\, 
%\biggl( 
[{\rm Li}_2(y)          \nonumber \\
    \zz  + 2{\rm Li}_2(-y) ]
%\biggr)  
%\nonumber \\ \zz 
+ 2 \log^2 y\, 
%\biggl( 
[\log(1-y) + 2 \log(1+y) ]
%\biggr)
\,.\nonumber 
\end{eqnarray}

\vskip 0.3cm

%\section{Discussions and Conclusions}

%Testing some reference as seen in \cite{epodd:LEB93} also in
%\cite{A-W:RRu88,tub:DKn89}. \smallsample

%\smallsample

%\section{Acknowledgments}
%\sample

Author
%(A. V. K.)
would like to express his sincerely thanks to the Organizing
Committee and 
%especially 
to Professors D.V. Shirkov and V.A. Ilyin 
for the possibility to present the poster
%kind invitation 
%and the financial support
at  such remarkable Conference.
%, and V.S. Fadin, L.L. Jenkovszky
%and L.N Lipatov 
%and E.?. Martynov
%for fruitful discussions.

\vskip -0.3cm

%%%%%%%%%%%%%%%%%%%%%%%%%%%%%%%%%%%%%%%%
% look into test.bbl for examples to manually generate references
% the stuff below interfaces BibTeX
%%%%%%%%%%%%%%%%%%%%%%%%%%%%%%%%%%%%%%%%
\nocite{*}
\bibliographystyle{aipproc}
\bibliography{ai}

%%%%%%%%%%%%%%%%%%%%%%%%%%%%%%%%%%%%%%%%
% only go to one column if you really need long tables of more than
% one page at the end of the document (see remarks in aipguide)
% otherwise you would end the document here with \end{document}
%%%%%%%%%%%%%%%%%%%%%%%%%%%%%%%%%%%%%%%%

\end{document}